\documentclass[a4paper,amsmath]{cas-sc}
\usepackage[authoryear,longnamesfirst]{natbib}

\usepackage{threeparttable}
\usepackage[section]{placeins}           %
\usepackage{threeparttable}
\usepackage{graphicx}                    % For eps figures, newer & more powerfull
\newcommand{\aap}{    {\it Astron. Astrophys.}}

\newcommand{\apj}{    {\it Astrophys. J.}}
\newcommand{\apjl}{   {\it Astrophys. J. Lett.}}

\newcommand{\grl}{    {\it Geophys. Res. Lett.}}
\newcommand{\prl}{    {\it Phys. Res. Lett.}}

\newcommand{\mnras}{  {\it Mon. Not. Roy. Astron. Soc.}}

\newcommand{\solphys}{{\it Solar Phys.}}
\newcommand{\ssr}{    {\it Space Sci. Rev.}}

\begin{document}
\let\WriteBookmarks\relax
\def\floatpagepagefraction{1}
\def\textpagefraction{.001}

\shorttitle{Comparisons of physics-based prediction}
\shortauthors{Jiang et al.}

\title[mode = title]{Comparison of physics-based prediction models of solar cycle 25}

\tnotemark[1]
%%%%%%%%%%%%%%%%%%%%%%%%%%%%%%%%%%%%%%%%%%%%%%%%%%%
% Author
\author[1,2]{Jie Jiang}[type=editor,
orcid=0000-0001-5002-0577]
\cormark[1]
\fnmark[1]
\ead{jiejiang@buaa.edu.cn}

\address[1]{School of Space and Environment, Beihang University, Beijing, China}
\address[2]{Key Laboratory of Space Environment Monitoring and Information Processing of MIIT, Beijing, China}

\author[1,2]{Zebin Zhang}

\author[3]{Krist\'{o}f Petrovay}
\address[3]{ELTE E\"{o}tv\"{o}s Lor\'{a}nd University, Department of Astronomy, Budapest, Hungary}

\cortext[cor1]{Corresponding author:Jie Jiang}

% \author[addressref={aff1,aff2},email={jiejiang@buaa.edu.cn}]{\fnm{Jie}~\lnm{Jiang}\orcid{0000-0001-5002-0577}}
% \author[addressref={aff1,aff2}]{\fnm{Zebin}~\lnm{Zhang}}
% \author[addressref={aff3}]{\fnm{Krist\'{o}f}~\lnm{Petrovay}}
% \address[id=aff1]{School of Space and Environment, Beihang University, Beijing, China}
% \address[id=aff2]{Key Laboratory of Space Environment Monitoring and Information Processing of MIIT, Beijing, China}
% \address[id=aff3]{ELTE E\"{o}tv\"{o}s Lor\'{a}nd University, Department of Astronomy, Budapest, Hungary}

%%% Abstract
\begin{abstract}[S U M M A R Y]
Physics-based solar cycle predictions provide an effective way to verify our understanding of the solar cycle. Before the start of cycle 25, several physics-based solar cycle predictions were developed. These predictions use flux transport dynamo (FTD) models, surface flux transport (SFT) models, or a combination of the two kinds of models. The common physics behind these predictions is that the surface poloidal fields around cycle minimum dominate the subsequent cycle strength. In the review, we first give short introductions to SFT and FTD models. Then we compare 7 physics-based prediction models from 4 aspects, which are what the predictor is, how to get the predictor, how to use the predictor, and what to predict. Finally, we demonstrate the large effect of assimilated magnetograms on predictions by two SFT numerical tests. We suggest that uncertainties in both initial magnetograms and sunspot emergence should be included in such physics-based predictions because of their large effects on the results. In addition, in the review we put emphasis on what we can learn from different predictions, rather than an assessment of prediction results.
\end{abstract}

%%%%%%%%%%%%%%%%%%%%%%%%%%%%%%%%%%%%%%%%%%%%%%%%%%%
%% Keywords
%
\begin{keywords}
Solar magnetic fields, Solar Cycle, Physics-based prediction, Uncertainties
\end{keywords}

\maketitle

%%%%%%%%%%%%%%%%%%%%%%%%%%%%%%%%%%%%%%%%%%%%%%%%%%%
%% Sections
%
%\iffalse
\section{Introduction}\label{s:Introduction}

The solar cycle is becoming one of the most important natural cycles \citep{Choudhuri2015} since the Sun's activity dominates Earth's space environment, which plays an essential role in our technological society. Solar cycles vary in both amplitude and length. As a consequence, solar cycle forecasting is becoming a necessity for some practical needs, e.g., the planning of space missions. Since there are still many outstanding questions concerning the physics behind the solar cycle, developing reliable solar cycle prediction schemes remains an important aspect of research in solar-terrestrial physics and astrophysics.

The last solar cycle, cycle 24, has been the weakest during the past century. This provides a rare opportunity to test the existing prediction methods. The reviews by \cite{Pesnell2008, Nandy2021} classify the methods into physics-based models, precursors, spectral, machine learning, statistical methods, and so on. The latter 3 methods can be categorised together as extrapolation methods, as suggested by \cite{Petrovay2020}. The unusually weak cycle 24 has confirmed that the precursor method based on the polar field (or axial dipole field) at cycle minimum has the highest predictive skill. Polar field measurements by direct observations (e.g., see Figure 2 of \cite{Jiang2007} and Figure 10 of \cite{Jiang2018}) and by geomagnetic proxies \citep{Ohl1979, Wang2009b} alike indicate that the polar field (or axial dipole field) at cycle minimum has a good correlation with the subsequent cycle strength. This correlation is essential in the context of the review. In the following, the default correlation denotes this one.

The correlation, on the one hand, adds a piece of evidence that the solar dynamo is the Babcock-Leighton (BL) type based on the notion that the surface sunspot emergence provides the poloidal seed field for the subsequent solar cycle. In contrast, fields and sunspots appearing on the surface are just a byproduct of the classical mean-field dynamo \citep{Parker1955, Steenbeck1966}. So in these models a tight correlation between the polar field at cycle minimum and the subsequent cycle strength might happen as presented by Figure 19 of \cite{Charbonneau2020}), but is not necessary. The polar field in the classical mean-field dynamo is not due to the decay of sunspots on the surface, and the polar field in the correlation results from the observable decay of sunspots. On the other hand, the correlation provides a constraint on the BL-type dynamo. The poloidal and toroidal field generation layers are spatially separated. The correlation constrains the transport time of the surface poloidal field to the layer of the toroidal field generation. The transport time should be less than a cycle. The physics-based predictions include the prediction schemes based on the BL-type dynamo model with data assimilation. The first dynamo-based solar cycle predictions appeared during cycle 23/24 minimum \citep{Dikpati2006a, Dikpati2006b, Choudhuri2007, Jiang2007}.

Solar cycle 25 started at the end of 2019 \citep{Nandy2020}. Several years before the cycle minimum, \cite{Cameron2016b} carried out a study to predict the polar field at cycle minimum, which was then used to predict the amplitude of cycle 25. This demonstrates that there is no need to wait until the cycle minimum to make the prediction. Independently, \cite{Hathaway2016} developed a similar method to predict the amplitude of cycle 25. Later further models with similar methods, e.g., \cite{Iijima2017, Jiang2018, Upton2018}, were suggested. In contrast to the polar field precursor method, which directly uses the observed correlation between the polar field at cycle minimum and the subsequent cycle strength \citep{Svalgaard2020}, the physics of the polar field generation is included in such kind of schemes by surface flux transport (SFT) models. Hence we also regard models of this type as physics-based predictions. Besides SFT-based predictions, dynamo-based cycle 25 predictions have also been developed, e.g., \cite{Bhowmik2018, Labonville2019, Guo2021}. Figure 4 of \cite{Labonville2019} and Figure 5 of \cite{Nandy2021} show the convergence in physics-based predictions of cycle 25 amplitudes, although some discrepancies still remain. Since there are several physics-based models available the question arises, what do they have in common, and what are the differences among these models? This review aims to compare some physics-based prediction models of cycle 25 from a methodological point of view.

The paper is organized as follows. A brief introduction to the physical models, namely SFT models and flux transport dynamo (FTD) models are given in Section 2. Section 3 presents a comparison of 7 physics-based prediction models of cycle 25. Section 4 focuses on a comparison of the effects of the magnetograms used as the initial condition on predictions. Section 5 concludes the paper.

\section{A brief introduction to the physical models used by physics-based predictions}\label{s:SFT_FTD}
\subsection{Surface flux transport models}\label{s:sft}
SFT models were initiated in 1980s \citep[e.g.][]{DeVore1984,Sheeley1985,DeVore1987,Wang1989}. The original idea can be traced back to \cite{Leighton1964}. The models describe the evolution of the large-scale magnetic field, which is assumed to be purely vertical $B_r$, over the solar surface. The sunspot group emergence provides the flux source $S(\theta,\phi,t)$. The flux of sunspot groups then evolves under the effects of a random walk due to supergranular flows, which are usually approximated by turbulent diffusion \citep{Leighton1964}, and the effects of large-scale surface flows including the meridional flows $\upsilon(\theta, \phi, t)\bf{e_\theta}$ and differential rotation $u(\theta, \phi, t)\bf{e_\phi}$. In most SFT models, both large-scale flows are taken as axisymmetric and time independent, namely, $\upsilon(\theta)\bf{e_\theta}$+$u(\theta)\bf{e_\phi}$ and $u(\theta)=R_\odot\sin\theta\Omega(\theta)$, where $\Omega(\theta)$ is the differential rotation. The governing equation is the radial component of the MHD induction equation at $r=R_\odot$ under the assumptions that $B_\theta|_{r=R_\odot}$=0, $B_\phi|_{r=R_\odot}$=0, $\frac{\partial B_\theta}{\partial r}|_{r=R_\odot}$=0 as follows:
\begin{eqnarray}
\frac{\partial B_r}{\partial t}=
     &-&\frac{1}{R_{\odot}\sin \theta} \frac{\partial } {\partial
     \phi}  \left(u B_r \right)
      -\frac{1}{R_{\odot}\sin \theta} \frac{\partial}{\partial \theta}
     \left(\upsilon
           B_r \sin\theta \right) \nonumber \\ \noalign{\vskip 2mm}
     &+& \eta_s\left[\frac{1}{R_{\odot}^2\sin \theta}
     \frac{\partial}{\partial \theta}
     \left(\sin \theta \frac{\partial B_r}{\partial \theta} \right)
         + \frac{1}{R_{\odot}^2\sin^2 \theta}\frac{\partial^2
     B_r}{\partial \phi^2}\right] \nonumber \\ \noalign{\vskip 2mm}
    &+&  S(\theta,\phi,t),
\label{eqn:SFT}
\end{eqnarray}
where $\theta$ and $\phi$ are heliographic colatitude and longitude, respectively. And $\eta_s$ is the uniform turbulent diffusivity.  Appendix of \cite{DeVore1984} provides another method to derive the SFT equation through spatially averaging the radial
component of the induction equation. Past studies have shown that SFT simulations are an effective way to understand the polar field evolution \citep[e.g.,][]{Wang2020}.

Two groups of parameters determine the evolution of the large-scale magnetic field $B_r(\theta, \phi, t)$, including the polar field. One is the transport parameters over the surface, including the strength of the turbulent diffusivity $\eta_s$ and profiles of the differential rotation $\Omega(\theta)$ and meridional flow $\upsilon(\theta)$. The other group of parameters is relevant to the flux source, namely, the magnetic field distribution of sunspot emergence.

There are fewer disagreements on the differential rotation owing to global helioseismology. But the turbulent diffusivity and meridional flow are less constrained by observations. Table 6.2 of \cite{Schrijver2000} summarizes various attempts to measure $\eta_s$ from observations. The values fall in the range 1.1-6.0 $\times$10$^{12}$ cm$^2$s$^{-1}$. \cite{Chae2008} show that the magnetic diffusivity originating from turbulence is determined by the characteristic length scale. For a typical size of a supergranule, a turbulent diffusivity $\eta_s$ is about 2.0$\times$10$^{12}$ cm$^2$s$^{-1}$ estimated based on the scaling relation of the turbulent cascade. Measurements of turbulent diffusivity in direct numerical simulations fall in the range 1.1-3.4$\times$10$^{12}$ cm$^2$s$^{-1}$ based on magnetic energy decay rates \citep{Cameron2011}, although some simulations are against treating the magnetic field as a passive scalar, e.g., \cite{Thaler2017} who regard that ``retraction'' below the surface driven by magnetic force is an effective mechanism of flux cancellation.

Instead of the diffusion approximation of the supergranular random walk, \cite{Upton2014} introduce a purely advective surface flux transport model. Supergranular flows are modeled explicitly by using vector spherical harmonics. A spectrum of spherical harmonics is used to create convection cells that reproduce the observed spectral characteristics. The spectral coefficients were evolved at each time step to give the cells finite lifetimes. The method introduces stochastic variations in flux transport.

The meridional flow shows divergent profiles based on three major methods, namely, feature tracking, direct Doppler measurements, and helioseismology. The divergence mainly results from the systematic effects in different methods \citep{Mahajan2021} and the time variation in the meridional flow itself within each solar cycle primarily due to the inflows toward active regions \citep{Chou2001,Cameron2010b}. \cite{Petrovay2020b} demonstrate that the parameter $\varPi=R_\odot\eta_s/\Delta_\upsilon$ plays an essential role in the SFT models, where
\begin{equation}
\Delta_\upsilon=\frac{d\upsilon}{d\lambda}\mid_{\lambda=0}.
\end{equation}
$\Delta_\upsilon$ corresponds to the divergence of the meridional flow at the equator. The parameter $\varPi$ shows the competition of flux transport between the poleward meridional flow and equatorward diffusion near the equator, which determines the cross-equatorial flux. Direct Doppler measurements \citep[e.g.,][]{Ulrich2010} and helioseismology \citep[e.g.,][]{Gizon2008,Gonzalez2008,Zhao2014,Liang2018} show that $\Delta_\upsilon$ falls in the range $\sim$0.8-$\sim$1.4 ms$^{-1}$deg$^{-1}$. The value depends on the solar cycle phase. It is larger at cycle maximum. Temporally averaged results based on the magnetic feature tracking \citep[e.g.,][]{Komm1993, Hathaway2011,Mahajan2021} show that $\Delta_\upsilon$ falls in the rage $\sim$0.4-$\sim$0.6 ms$^{-1}$deg$^{-1}$. Recently \cite{Mahajan2021} show $\Delta_\upsilon$=$\sim$0.95 based on an improved  magnetic feature tracking method. Figure 3 in \cite{Petrovay2020b} presents 4 profiles used by SFT models. The profile used by \cite{Wang2017}, namely, $13\tanh(|\lambda|/6^\circ)\cos^2\lambda$, has the $\Delta_\upsilon$ value of 2.17 ms$^{-1}$deg$^{-1}$, which is much larger than all of the available meridional flow measurements. In most other SFT simulations, $\Delta_\upsilon$ is close to 0.5 ms$^{-1}$deg$^{-1}$. A collection of different $\Delta_\upsilon$ values derived by observations and used by SFT simulations is presented in Table \ref{table:1}. Please note that if the analytical profile of the flow from observations is not given in a publication, the $\Delta_\upsilon$ value is measured based on the corresponding figure. Such values have limited accuracy and hence the symbol ``$\sim$'' is used in Table \ref{table:1}.

\begin{table}[tp]
\caption{A collection of different $\Delta_\upsilon$ values derived by observations and used by SFT simulations. } % title of Table
\label{table:1} % is used to refer this table in the text
\centering % used for centering table
\begin{tabular}{p{4cm}p{4.5cm}p{3.5cm}} % centered columns (4 columns)
\hline\hline % inserts double horizontal lines
 Reference     & Method &   $\Delta_\upsilon$ value (ms$^{-1}$deg$^{-1}$)\\ % table heading
\hline % inserts single horizontal line
Figure 1b of \cite{Gizon2008} & Time-distance helioseismology, MDI data, near the solar surface, 1996-2002 &  $\sim$1.15 \\
Figure 3a of \cite{Zhao2014} & Time-distance helioseismology, HMI data, a depth of 0-1 Mm, 2010-2012  &  $\sim$0.8-$\sim$1.23 \\
Figure 8f of \cite{Liang2018} & Time-distance helioseismology, MDI \& HMI data, at the surface, 1996-2017 &  1.21\\
Figure 3c of \cite{Gizon2020}   & Time-distance helioseismology, MDI and GONG data, at the surface, 1996-2019 & $\sim$1.0-$\sim$1.2\\
Figure 1a of \cite{Gonzalez2008} & Ring-diagram analysis, GONG data, at 3.1Mm, 2001-2006 &  $\sim$0.8-$\sim$1.4 \\
Figure 7 of \cite{Ulrich2010} & Doppler shift, MWO data, Fe I spectral line at 5250{\AA}, averages over the full cycles 22 and 23 &  $\sim$1.2  \\
Equation (1b) and Table 1b of \cite{Komm1993} & Magnetic feature tracking, NSO magnetograms, 1978-1990 & [0.42, 0.66]  \\
Equation (9) of \cite{Hathaway2011} & Small magnetic elements tracking, MDI magnetograms, 1996-2010 &   0.51 \\
Figure 6 of \cite{Mahajan2021} & Magnetic feature tracking, MDI \& HMI magnetograms, 1996-2020 & $\sim$0.95\\
\hline
\cite{Wang2009} & Applied in SFT simulations & $\infty$\\
\cite{Wang2017} & Applied in SFT simulations & 2.17\\
\cite{Ballegooijen98,Cameron2016b, Bhowmik2018, Jiang2018} & Applied in SFT simulations & 0.46\\
\cite{Whitbread2018} & From optimisation, applied in SFT simulations & 0.43\\
\hline\hline
\end{tabular}
\end{table}

The distribution of flux source was usually treated by bipolar magnetic regions (BMRs) approximation in SFT models. The BMR approximation means that an AR is simplified as a leading and following bipolar with symmetric configuration but opposite polarities. The area, location (latitude and longitude), and tilt angle are the key parameters to determine flux distribution. The number of BMRs is proportional to the cycle strength.  The tilt angle is usually derived based on Joy's law. The traditional way to deal with the parameters makes the  SFT process a linear one, which means that the generation of the polar field at cycle minimum is proportional to the cycle strength. This is not consistent with observations, see Figure 3 of \cite{Jiang2007} and Figure 5 of \cite{Munoz-Jaramillo2013}. Studies during the past decade have revealed that both nonlinear and stochastic mechanisms resulting from the flux source parameters are involved in the polar field evolution.

\cite{Jiang2014} determine the tilt scatter using the observed tilt angle data and study effects of this scatter on the evolution of the solar surface field using SFT simulations. The results show that the uncertainty of the axial dipole field caused by the tilt scatter is over 30\% of the average strength. The sunspot groups which emerge closer to the equator have a larger contribution to the axial dipole strength than that emerge at higher latitudes because the lower latitude sunspot emergence tends to contribute more net cross-equatorial flux. The latitudinal dependence of sunspot groups' contribution to the axial dipole strength obeys a Gaussian profile. The result is further demonstrated by \cite{Nagy2017} and \cite{Whitbread2018}. \cite{Jiang2019} for the first time demonstrate that the evolution of a complex AR, i.e., $\delta$-spot like AR12693, could be significantly different from the (tilt) BMR approximation. An initial strongly positive dipole moment could end up with a strongly negative value. The importance to include the realistic configuration of complex sunspot groups was further verified by \cite{Yeates2020} and \cite{Wang2021}. The difference between the realistic configuration and the BMR approximation results from the mechanism mentioned above. That is, the flux at lower latitudes contributes more to the axial dipole strength than that at higher latitudes since the cross-equatorial flux determines the contribution to the axial dipole field.

Beside the stochastic property, sunspot emergence shows systematic properties that stronger cycles tend to have higher mean latitudes \citep{Li2003, Solanki2008, Jiang2011} and lower tilt angle coefficients, which exclude the latitudinal dependence of the tilt angles \citep{Dasi2010, Jiao2021}. \cite{Jiang2020} demonstrates that the systematic change in latitude and tilt plays the same important role in modulating the polar field generation. By including both forms of nonlinearities resulting from the cycle-dependent mean latitudes and tilt angle coefficients of sunspot groups as the source term in SFT simulations for multiple cycles, \cite{Cameron2010} and \cite{Bhowmik2018} obtained the regular polar field reversal. This provides a piece of evidence that the additional term of the radial diffusion proposed by \cite{Schrijver2002} is not required in the SFT process.

In addition, \cite{Cameron2010} and \cite{Bhowmik2018} multiplied a factor of 0.7 to the tilt to mimic the effects of the near-surface inflows toward sunspot groups when they got the regular polar field reversal. \cite{Martin-Belda2017} show that inflows towards active regions are a potential non-linear mechanism modulating the solar cycle. Inflows reduce the axial dipole strength at the end of the cycle by ~30\% with respect to the case without inflows in cycles of moderate activity \citep{Jiang2010}. This ratio varies by ~9\% from very weak cycles to very strong cycles. Inflows might be responsible for the anti-correlation between the tilt coefficient and the cycle strength. So in SFT simulations of multiple cycles inflows might not be required if the cycle-dependent mean tilt angle coefficients of sunspot groups are included.

For the initial magnetic field distribution at the start of the simulations, the analytic profile of the surface large-scale field distribution is usually used when simulations start from cycle minima. The observed synoptic maps are used for the cycle predictions. We will demonstrate that the low latitude flux distribution in magnetograms has significant effects on the surface flux evolution in Section \ref{sec:detailedCom}.

\subsection{Flux transport dynamo models}\label{s:ftd}

The success of the SFT models in 1980s led to the birth of the FTD models in 1990s \citep{Wang1991, Choudhuri1995, Durney1995}. During the past decades FTD models are the workhorse of the understanding of the solar cycle. The equations for the axisymmetric model using spherical coordinates ($r, \theta, \phi$) are
\begin{equation}
\label{eqn:ftdA}
 \frac{\partial A}{\partial t}+\frac{1}{r\sin \theta}(\mathbf{v_p}\cdot\nabla)(r\sin\theta A)=\eta\left(\nabla^{2}-\frac{1}{r^{2}\sin^{2}\theta}\right) A+\alpha B
\end{equation}
\begin{eqnarray}
\label{eqn:ftdB}
\frac{\partial B_\phi}{\partial t}+\frac{1}{r}\left[\frac{\partial}{\partial
r}(rv_rB_\phi)+\frac{\partial}{\partial\theta}(v_\theta B_\phi)\right]
=\eta\left(\nabla^2-\frac{1}{r^2\sin^{2}\theta}\right)B_\phi\nonumber\\
+r\sin\theta(\textbf{B}_p\cdot\nabla)\Omega+\frac{1}{r}\frac{d\eta}{dr}\frac{\partial}{\partial r}(rB_\phi),
\end{eqnarray}
where $B_\phi(r,\theta,t)\mathbf{e_\phi}$ and $\mathbf{B}_{p}=\nabla\times[A(r,\theta,t)\mathbf{e_\phi}]$ correspond to the toroidal and poloidal magnetic field, respectively. The large-scale flow fields include the meridional flow $\mathbf{v_p}=\upsilon_r\mathbf{e_r}+\upsilon_\theta\mathbf{e_\theta}$ and differential rotation $\Omega\mathbf{e_\phi}$, usually both of which are time independent in FTD models.

The $\alpha$-term of Eq.(\ref{eqn:ftdA}) is the poloidal field source term, which corresponds to the simplified profile of the BL process. It describes the emergence and decay of the tilt sunspot groups over the solar surface. Since the term is observable, it provides the possibility for the models to be applied in the solar cycle prediction. Some FTD dynamo models include an additional $\alpha$-term working at the base or the bulk of the convection zone for two possible reasons. One is that some FTD models prefer the quadrupolar solution \citep[e.g.,][]{Dikpati1999}. The additional $\alpha$-term helps to get the solar-like dipolar solution \citep[e.g.,][]{Dikpati2001}. Another group \citep{Passos2014,Hazra2020} argued that the BL-type FTD models are not self-excited. The additional poloidal field source effective on weak fields is necessary for the self-consistent recovery of the sunspot cycle from the Maunder-like grand minima. \cite{Xu1982,Eddy1983,Wang2019,Arlt2020} show that there were naked-eye sunspots recorded through the whole Maunder minimum. The naked-eye sunspots indicate that there were sunspots, even big spots, emerged on the Sun through the Maunder minimum. Moreover, the additional $\alpha$-term could be physically unavoidable, as in the low Rossby number environment of the deep convection zone, any up- or downflow tends to acquire a swirl and convert advected poloidal and toroidal fields into each other. But they might not contribute to the net toroidal flux in each hemisphere for sunspot emergence since \cite{Cameron2015} demonstrated that the dipole field generated by the decay of active region is sufficient to generate the toroidal flux emerging in the course of the cycle. Hence the additional $\alpha$-term in a BL-type FTD model might not be required.

In most FTD models, the toroidal field is generated in the tachocline, so that the generation layers of the poloidal and toroidal fields are spatially segregated. The transport parameters, namely, the meridional flow $\mathbf{v_p}$ and turbulent diffusion $\eta$ are required by the FTD models. The time delay provides the other advantage of the FTD models to be applied in the cycle prediction. Recently, \cite{Zhang2022} develop a new BL-type solar dynamo operating in the bulk of the convection zone to meet the requirement in the progress in understanding fully convective late-M dwarfs.

The main constraint on the turbulent diffusivity is given by the mixing length theory, which shows that the diffusivity is in order of 10$^{13}$ cm$^2$s$^{-1}$. With such strong diffusivity, FTD simulations cannot yield sustainable magnetic cycles \citep{Munoz-Jaramillo2011}. Hence different assumptions on the strength of turbulent diffusivity in the bulk of the convection zone were made in different models. The assumed values are usually much lower than the expected value. The diffusivity used in the prediction of \cite{Dikpati2006a} and \cite{Dikpati2006b} is in the range 10$^{10}$-10$^{11}$ cm$^2$s$^{-1}$. The prediction model of \cite{Choudhuri2007} and \cite{Jiang2007} used different diffusivities for the poloidal and toroidal field, which are 2.4$\times$10$^{12}$ cm$^2$s$^{-1}$ and 4$\times$10$^{10}$ cm$^2$s$^{-1}$, respectively. The dynamo model was based on \cite{Chatterjee2004}, who argue that within the main body of the convection zone, the action of turbulent diffusivity on the strong toroidal field is considerably suppressed. The strong diffusivity for the poloidal field also accounts for a solar-like dipolar solution in such kinds of models since the strong diffusivity promotes the couple of two hemispheres.

In the FTD models, the meridional flow plays an essential role. Its strength dominates the cycle period. And its equatorward branch is regarded to be responsible for the equatorward migration of the toroidal field. In Section \ref{s:ftd} we show the divergences in the surface poleward flows measurements. Its internal structures including the penetration depth, the location of the return flow and so on, are more controversial \citep{Zhao2013,Schad2013,Rajaguru2015,Gizon2020}. The meridional flow combined with the turbulent diffusion determines the so-called magnetic memory, which means how many of the following cycles show the effect of the current polar field. The different assumptions of the two transport parameters lead to different magnetic memories \citep{Yeates2008}. A strong turbulent diffusivity for the poloidal field \citep[e.g.,][]{Chatterjee2004} or a fast equatorward meridional flow ($>\sim4.5$ms$^{-1}$) at the toroidal field location \citep[e.g.,][]{Lemerle2017} can lead to the correlation between the polar field and the subsequent cycle strength, which is consistent with the observations. Recently \cite{Kumar2021b} found that the memory of the polar field changes from multiple to one cycle with the increase of the supercriticality of the dynamo, and the memory is independent of the importance of various turbulent transport processes in FTD models.
%\fi

\section{Comparison among 7 physics-based prediction models of cycle 25} \label{s:Comparisons}
\subsection{Main ingredients of the prediction models}

\begin{figure}[htbp]
\centering
\includegraphics[scale=0.25]{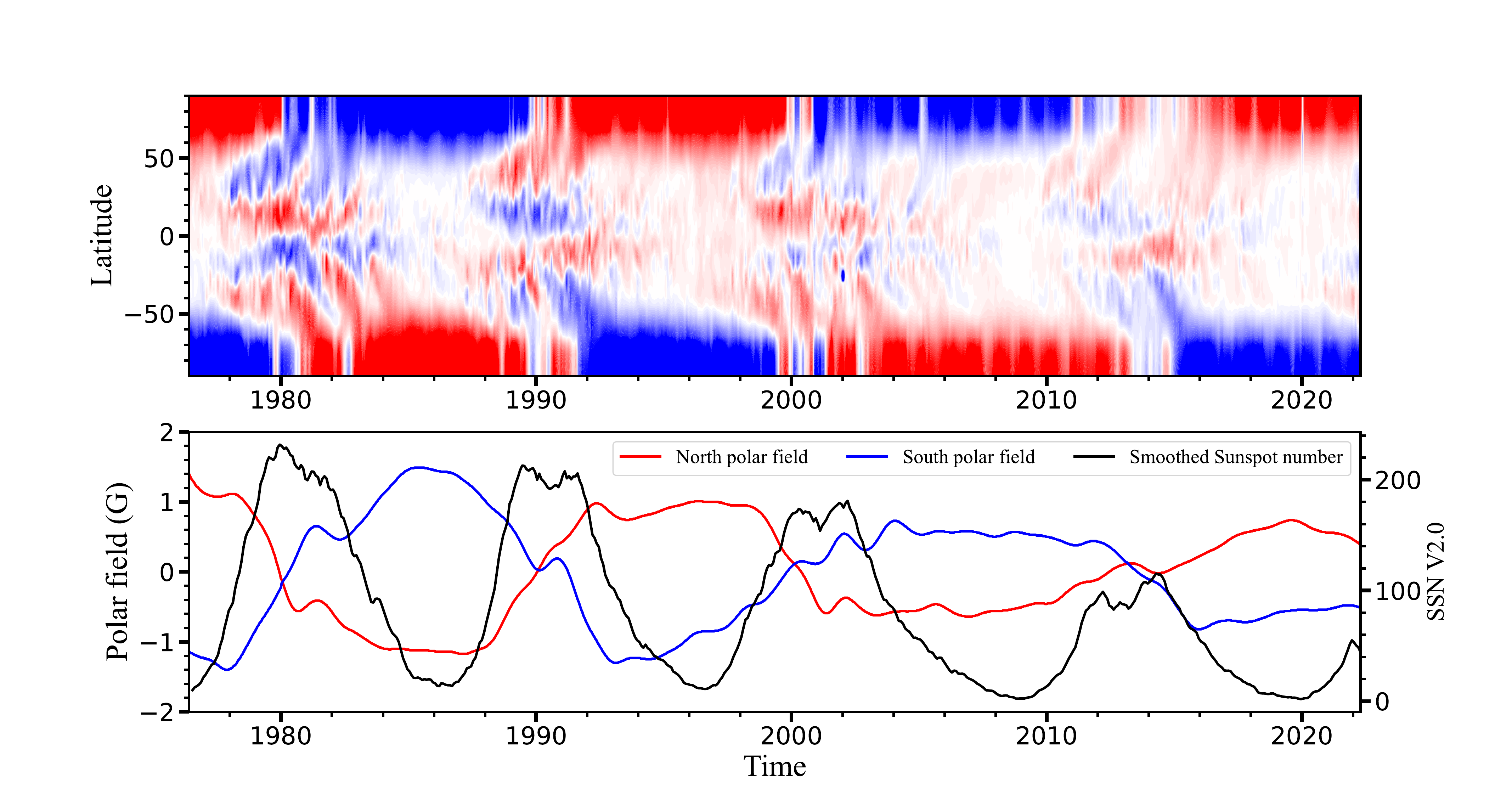}
\caption{Time evolution of the longitudinal averaged photospheric magnetic field (upper panel) from WSO, sunspot number (black curve), northern (red curve, lower panel) and southern polar field (blue curve, lower panel) from WSO.}
\label{fig:WSO}
\end{figure}

In all physics-based cycle 25 predictions that we will compare, the surface sunspot emergence provides the poloidal seed field for the future cycle, which means that all models work in the framework of the BL-type dynamo. The surface poloidal field, which is assumed to be purely radial, $B_r(R_\odot, \theta, \phi, t)$, is the key observable parameter to determine the next cycle strength. With the assumption of the axisymmetric property of $B_r$, the time evolution of the longitudinal averaged photospheric synoptic magnetograms, $\left\langle B_r(R_\odot, \theta, t)\right\rangle$, is the data which have the most direct connection with the physics-based prediction models. Upper panel of Figure \ref{fig:WSO} shows $\left\langle B_r(R_\odot, \theta, t)\right\rangle$ from 1976 to the present based on the Wilcox Solar Observatory (WSO) \footnote{http://wso.stanford.edu/}. The polar fields are extrapolated based on the method shown in \cite{Guo2021} Hereafter we refer to the quantity relevant to observations and used by the physics-based prediction models as \textit{the predictor}. Since observations have shown that the field around the minima of the activity cycle is essential, some models only consider the surface large-scale field at cycle minimum, $\left\langle B_r(R_\odot, \theta, t_{min})\right\rangle$. Some models only consider the polar component of the poloidal field, namely, the polar field. Since the definition of the polar cap is a little bit arbitrary, the axial dipole strength $D$(t), which is defined as
\begin{equation}
\label{eqn:dm_t}
D(t)=\frac{3}{2}\int_0^{\pi}\left\langle B_r\right\rangle(R_\odot, \theta, t)\cos\theta\sin\theta d\theta,
\end{equation}
is also frequently used. In the literate $D$(t) is usually referred as the axial dipole moment \citep[e.g.,][]{Wang1991}. It actually is the amplitude of the coefficient of the $Y_1^{0}$ term in a spherical harmonic expansion of $B_r(R_\odot, \theta, \phi, t)$ \citep{Petrovay2020}. The first ingredient that contributes to the difference among different prediction models are the four predictors mentioned above, i.e., the observed $\left\langle B_r(R_\odot, \theta, t)\right\rangle$, $\left\langle B_r(R_\odot, \theta, t_{min})\right\rangle$, polar field at cycle minimum, and axial dipole strength at cycle minimum.

%\iffalse
\begin{figure}[t]
  \centering
  \includegraphics[scale=0.5]{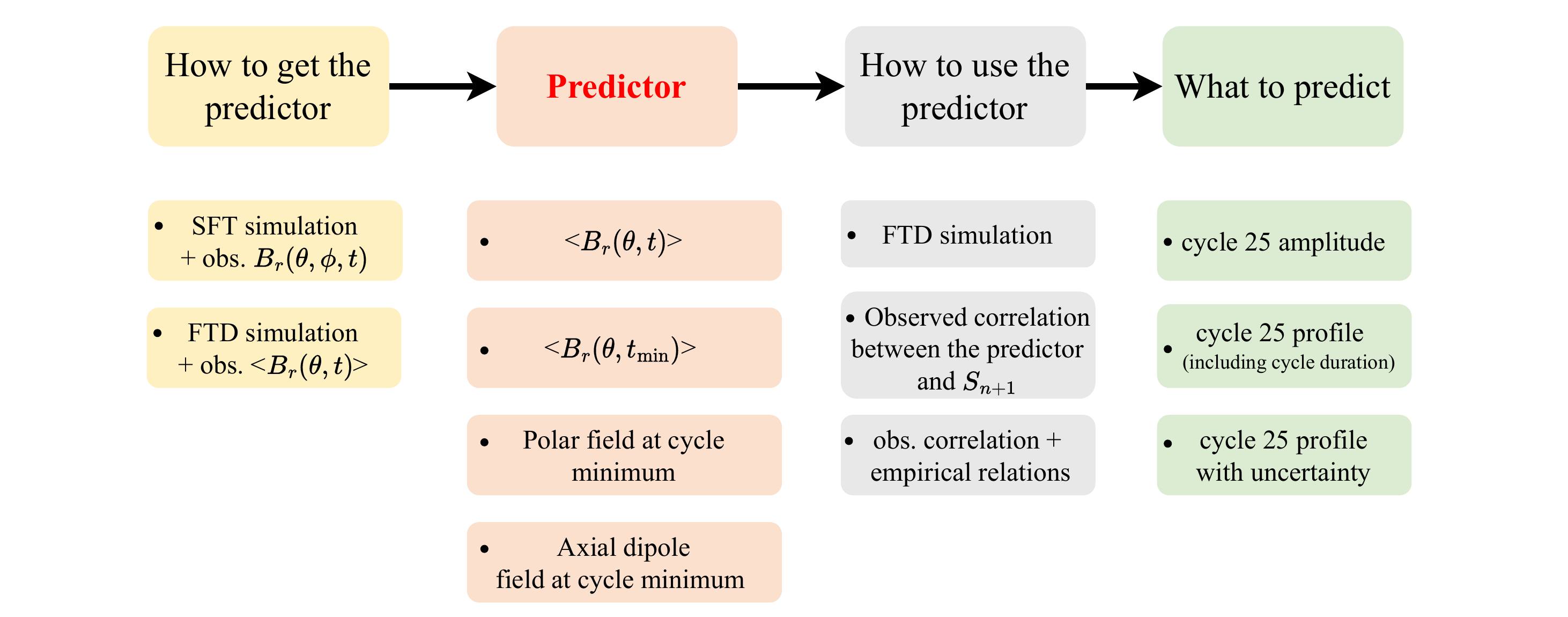}
  \caption{Four key ingredients of the physics-based prediction models (top line) and different methods to derive each ingredient in the physics-based predictions.}
  \label{fig:skeleton}
  \end{figure}

  \begin{figure}[h!]
  \centering
  \includegraphics[scale=0.5]{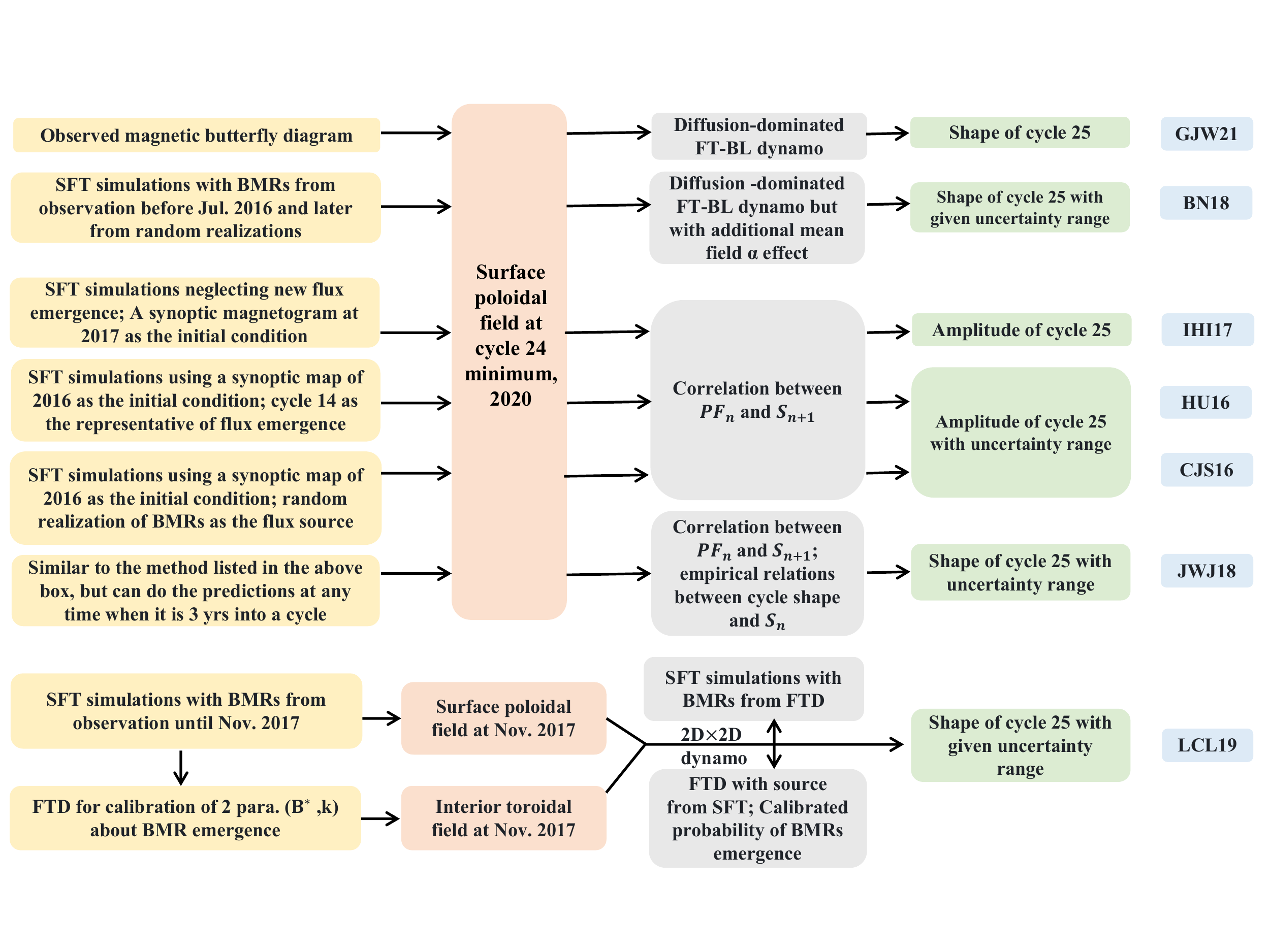}
  \caption{Comparisons of different methods used by 7 physics-based cycle 25 predictions.}
  \label{fig:diffMethods}
  \end{figure}
  %\fi

Besides the different predictors, prediction models differ from each other in the following three ingredients: (1) how to get the predictor, (2) how to apply the predictor in the prediction, and (3) what to predict.

SFT simulations provide an effective way to get the predictor, e.g., polar field or axial dipole strength at cycle minimum. The direct or indirect observed synoptic or synchronic map $B_r(R_\odot, \theta, \phi, t_0)$ at a time $t_0$ before the cycle 24/25 minimum is used as the initial condition. After getting the polar field (or axial dipole strength, $PF_n$) at the cycle 24/25 minimum, the correlation between it and the subsequent cycle strength ($S_{n+1}$) can be used to predict the subsequent cycle strength. But the profile of the cycle cannot be derived based on the correlation. Since empirical relations between the cycle amplitude and cycle shape are available, some prediction models further apply the empirical relations to predict the shape of the subsequent cycle, even including uncertain ranges at different phases of cycle 25.

FTD models provide another effective way to get the predictor. Some models continuously assimilate the $\left\langle B_r(R_\odot, \theta, t)\right\rangle$ into FTD simulations until the prediction timing. Some models just use the observed $\left\langle B_r(R_\odot, \theta, t_{min})\right\rangle$ as the initial condition. The corresponding toroidal field from FTD models gives the prediction of the cycle 25 profile, including the cycle duration. But usually the predicted cycle period does not change with the cycle amplitude since dynamo models have fixed cycle period.

Figure \ref{fig:skeleton} summarizes the four key ingredients of physics-based predictions and different methods to derive each ingredient. In the next subsection, we concentrate on comparisons and analysis of the differences among the 7 available physics-based predictions of cycle 25 especially based on the four ingredients. The 7 models are: (1) \cite{Cameron2016b}, hereafter CJS16, (2) \cite{Hathaway2016}, hereafter HU16,  (3) \cite{Iijima2017}, hereafter IHI17,  (4) \cite{Jiang2018}, hereafter JWJ18,  (5) \cite{Bhowmik2018}, hereafter BN18, (6) \cite{Labonville2019}, hereafter LCL19, (7) \cite{Guo2021}, hereafter GJW21. A summary and comparison of the 7 models is presented in Figure \ref{fig:diffMethods}.

\subsection{The seven prediction models}
%\iffalse
\subsubsection{GJW21}
To verify whether the success of the dynamo-based prediction model developed for cycle 24 \citep{Jiang2007} is repeatable, GJW21 apply the method to the cycle 25 prediction. The same diffusion-dominated FTD model developed by \cite{Chatterjee2004} is utilized. The predictor in the prediction model is the observed WSO longitudinal averaged photospheric synoptic magnetograms at cycle minimum, $\left\langle B_r(R_\odot, \theta, t_{min})\right\rangle$. Since cycle 23 has the average cycle amplitude based on sunspot records since 1874, cycle 23 is used as reference to calibrate the model. That is, the averaged $\left\langle B_r(R_\odot, \theta, t)\right\rangle$ of three-year intervals before the cycle 22/23 minimum is regarded as the average strength corresponding to the result in that phase of the dynamo. The vector potential $A$ in Eq.(\ref{eqn:ftdA}) is derived based on $\left\langle B_r(R_\odot, \theta, t_{min})\right\rangle$. Its values at the cycle 22/23 minimum are used to calibrate the toroidal field strength of the dynamo model.

During every cycle minimum, the dynamo simulation is corrected based on values of $\left\langle B_r(R_\odot, \theta, t_{min})\right\rangle $ for every grid point above 0.85$R_\odot$ and then is resumed without interruption until the next cycle minimum. When the toroidal field at the tachocline exceeds a critical value, a corresponding sunspot emergence at the surface is given, so that the time evolution of the whole cycle 25 could be predicted by this method. The predicted rising phase and amplitude vary with the assimilated poloidal field strength. The decline phase and the cycle period do not depend on the input. Hence, although the prediction can give the profile of the cycle, the actual focus of the prediction model is just the cycle amplitude.

\subsubsection{CJS16}
To apply method used by GJW21, one must wait until the cycle 24/25 minimum to carry out the predictions using the observed data at the cycle 24/25 minimum. Since SFT models can reproduce the evolution of the solar surface magnetic field well, CJS16 make the first predict the polar field at the cycle 24/25 minimum with the SFT model about 4 years before the cycle 24/25 minimum. With this method, the prediction window can be extended to a few years before a minimum. The axial dipole strength is the predictor of the method. SFT models are used to get the predictor and then the correlation between the predictor and the subsequent cycle strength is used to estimate the cycle 25 strength. The method can only predict the range of cycle 25 strength, the uncertainties of which mainly result from the intrinsic randomness in sunspot emergence and in the assimilated magnetograms.

There are two key ingredients of the SFT simulations. One is the initial condition from observations for the SFT modeling. The other is to predict the BMR emergence during the remaining time of cycle 24 and then assimilate it as the flux source into the SFT model. The initial field of CJS16 is the observed synoptic magnetograms from SDO/HMI. The error due to the observed magnetograms is included in the final error estimates. CJS16 synthesize the BMR emergence based on the statistical relations given by \cite{Jiang2011} using 50 random realizations, so that the random components in number, tilt, latitude, and area of emerging BMRs are incorporated. Thus the uncertainty of the axial dipole strength at the end of the cycle minimum can be estimated.

\subsubsection{HU16}
HU16 also use an SFT model to predict the Sun's axial dipole strength at the cycle 24/25 minimum and further to give a prediction of cycle 25 amplitude based on the correlation. But there are some differences in details in their methods from that of CJS16.

%As for the transport parameters of the two SFT models, the random walk of supergranulations is modeled explicitly by HU16.  See Section \ref{s:sft} for more details. What CJS16 used is the turbulent diffusivity approximation. Other transport parameters are more or less identical.

CJS16 and HU16 used HMI synoptic magnetograms and synchronic maps as the SFT initial condition, respectively. The synchronic maps are constructed by using evolving supergranules and the observed axisymmetric flows to transport flux with data assimilated from HMI at 60 min intervals \citep{Upton2014}. The synchronic maps could have some differences, especially around the equator, from HMI synoptic map at a given time. Since we have shown that the low latitude flux plays an essential role in the polar field evolution, using different maps can result in a significant difference in the prediction of the polar field at the cycle minimum. This will be demonstrated in Section \ref{sec:detailedCom}.

The second difference between CJS16 and HU16 is the representation of the BMR emergence during the remaining time of the ongoing cycle. HU16 used cycle 14 as a representation of BMRs that would appear in 2017-2020. Only randomness in BMRs' tilt was considered. Although solar cycle shapes seem to form a family of curves well characterized by a single parameter, i.e., cycle strength $S_n$ \citep{Waldmeier1955}, the random processes involved in the flux emergence cause the scatter distributions of BMRs in their areas, latitudes, and tilts. An occasional emergence of big spots near the equator with large tilts could have significant effects on the polar field \citep{Jiang2014,Nagy2017, Wang2021} as we have mentioned in Section \ref{s:sft}. In contrast, CJS16 synthesized the BMR emergence based on the statistical relations using random realizations.

The third difference between CJS16 and HU16 is the way to deal with the convective motions and meridional flow. CJS16 did not investigate effects of the variation in the two ingredients on the results, and they used the constant turbulent diffusivity approximation of the convective motions. HU16 incorporated the variations in the convection pattern in their model by eight different realizations based on the changes in the spectral coefficients produced by the axismmetric flows. Furthermore, HU16 also included the cycle-related variation in the meridional flow. But their results show that the variation in the axial dipole strength caused by the Joy's law variation is large enough to overpower the variation due to changes in the meridional flow and the convective motions.

\subsubsection{IHI17}
IHI17 suggest that the axial dipole strength becomes approximately constant several years before each cycle minimum based on analysis of photospheric synoptic magnetograms of cycles 21-23. They further provide a piece of evidence that the time variation of the observed axial dipole strength agrees well with that predicted by their SFT model without introducing new emergence of magnetic flux. Hence they took a synoptic magnetogram in 2017 as the initial condition for an SFT simulation neglecting new flux emergence to predict the axial dipole strength in 2020. They suggest that the amplitude of cycle 25 is even weaker than cycle 24 based on the predicted axial dipole strength. As a whole, a major difference of the method from CJS16 is that the contribution of the sunspot emergence during the last few years of a cycle to the axial dipole strength at the cycle minimum is ignored by IHI17.

Based on Figure 9 of \cite{Jiang2018} and our latest update, the dipole moment changes $\Delta D$ for the last 3 years of cycles 21-24 minima are 0.01 G, -0.47 G, -0.41 G, and 0.08 G, respectively. They correspond to the difference between axial dipole strength at a cycle minimum $D(t_{min})$ and that of three years before a minimum $D(t_{min-3yr})$, i.e.,  $\Delta D$=$D(t_{min})$-$D(t_{min-3yr})$. The relative changes are 0.0\%, -13.3\%, -23.4\%,  and 4.73\%, which are not large variations. However, the 4 cycles only correspond to 4 random realizations of magnetic flux emergence. Figures 3g, 3h, and 3i of \cite{Jiang2020} show the time evolution of axial dipole strength of 3 sets of synthesized cycles with amplitudes of 70, 180, and 280, respectively. Each of them represents a typical weak, medium, and strong cycle. For each case, there are 100 random realizations. For the weak cycle with the cycle amplitude of 70, the maximum increase and decrease of dipole moment during the last 3 years are 2.28 G and -0.21 G. The average value and the standard deviation over 100 realizations are 1.12$\pm$0.52 G. For the medium cycle with the cycle amplitude of 180, the maximum increase and decrease are 2.3 G and -0.22 G. The average value and the standard deviation are 0.8$\pm$0.47 G. For the strong cycle with the cycle amplitude of 280, the maximum increase and decrease are 1.55 G and -0.61 G. The average value and the standard deviation are 0.29$\pm$0.48 G. An average dipole moment at cycle minimum is about 3.5 G. Hence the variation of the dipole moment due to the flux emergence during the last 3 years of a cycle can be above 60\% of the value at cycle minimum, especially for a weak cycle. The results of the observed cycles, cycles 21-24, are consistent with the property of a large number of synthesized cycles. A summary of the dipole moment changes for the last 3 years of observed and synthesized cycles is presented in Table \ref{table:2}.

The synthesized cycles indicate that flux emergence 3 years before cycle minimum could have large effects on the axial dipole strength at cycle minimum. They cannot be neglected when we make the cycle prediction before a cycle minimum. Moreover, past attempts have shown that the polar field precursor methods have large uncertainties and could fail \citep[e.g.,][]{Schatten1993} if used too early compared to cycle minimum \citep{Svalgaard2005}.

\begin{table}
\caption{A summary of the axial dipole strength changes for the last 3 years of observed and synthesized cycles.} % title of Table
\label{table:2} % is used to refer this table in the text
\centering % used for centering table
\begin{threeparttable}
\begin{tabular}{p{4.5cm}p{4.0cm}p{3.5cm}}
\hline\hline % inserts double horizontal lines
 Cycle     & Dipole strength change ($\Delta D$, in G) &   Percentage of the change\\ % table heading
\hline % inserts single horizontal line
  Observed cycle 21, $S_n=$232        &  0.01     &  0.0\%  \\
  Observed cycle 22, $S_n=$213        & -0.47    &  -13.3\% \\
  Observed cycle 23, $S_n=$181        & -0.41    &  -23.4\% \\
  Observed cycle 24, $S_n=$115        &  0.08    &   4.73\% \\
  Synthesized cycles, $S_n=$280 & 0.29$\pm$0.48 & -\\
  Synthesized cycles, $S_n=$180 & 0.80$\pm$0.47 & -\\
  Synthesized cycles, $S_n=$70 & 1.12$\pm$0.52 & -\\
 \hline \hline
\end{tabular}
      \begin{tablenotes}
        \footnotesize
         \item[1] $\Delta D$=$D(t_{min})$-$D(t_{min-3yr})$, where $t_{min}$ is a cycle minimum.
         \item[2] The third column is calculated by $\Delta D$/$D(t_{min})$.
         \item[3] $S_n$ is the maximum value of the monthly averaged sunspot number.
         \item[4] The synthesized cycles are from Figure 3 of \cite{Jiang2020}. The average value and standard deviation are calculated based on 100 realizations for each cycle strength.
      \end{tablenotes}
  \end{threeparttable}
\end{table}

\subsubsection{JWJ18}
JWJ18 also predict the subsequent cycle strength by predicting the dipole moment at cycle minimum using an SFT model. But JWJ18 can make the prediction at any phase of a cycle when it has been ongoing for more than three years. Furthermore, JWJ18 put emphasis on investigating the stochastic mechanisms and assessing their effects on the uncertainty of the prediction. They can predict not only the cycle strength with its uncertainty but also the possible shape of the future cycle. The shape of a solar cycle, i.e., the time evolution of the sunspot number, has systematic and random components. The cycle strength $S_n$ determines the systematic component of solar cycle shapes including the rising and maximum phases. As for the decline phase, all cycles decline in the same way on average as suggested by \cite{Cameron2016}. The random component depends on the cycle phase. During the maximum phase, the relative uncertainty is small. The properties of sunspot emergence, e.g., area, tilt, and latitude are also cycle-amplitude dependent. They also have random components, which are measured based on RGO sunspot area data set and Kodaikanal and Mount Wilson sunspot tilt angle data sets \citep{Jiang2011}. These random components significantly affect the predictability of the solar cycle. With these empirical relations, they can predict the solar cycle profile over one cycle with uncertainties.

\subsubsection{BN18}
BN18 employ both the SFT model and the FTD model for the prediction. They first use an SFT model to get the radial field evolution over the solar surface $\left\langle B_r(R_\odot, \theta, t)\right\rangle$ from 1913 to 2016. After that, they synthesize the sunspot emergence from 2016 (when the work was done) to 2020 (the cycle 24 minimum) to get the possible polar field at the cycle 24 minimum. Then at each cycle minimum they assimilate the surface radial field into their FTD model. Finally they give a prediction of the yearly mean sunspot number at the maximum of solar cycle 25. It is 118 with a predicted range of 109-139.

To get the century-scale radial field evolution they use SFT simulations. The parameters of their SFT model and the way to deal with the flux source term are similar to what \cite{Cameron2010} did. A major difference between the two studies is whether the simulations could reproduce observed cycle variability during cycles 21-24. The period of SFT simulations done by \cite{Cameron2010} is 1913-1976 using the RGO sunspot area dataset. Beyond 1976, the sunspot area dataset was shifted from RGO to USAF/NOAA. A constant factor of 1.4 is suggested by the website \footnote{http://solarcyclescience.com/activeregions.html} to maintain the consistency in the area measurements from the two data sources. \cite{Cameron2010} could not reproduce the surface poloidal field evolution consistent with observations from 1976 onwards with the 1.4 factor. Hence they limit their study in the range of 1913-1976. In contrast, \cite{Bhowmik2018} only multiply the factor of 1.4 to any active region area belonging to the USAF/NOAA database if its area is smaller than 206 micro-hemispheres. Thus they obtain the continuous large-scale field evolution from 1913 to 2016.

The dynamo model used by BN18 includes both surface $\alpha$-effect and classical mean-field $\alpha$-effect in the convection zone as an essential means for reproducing important observational features \citep{Passos2014}. \cite{Hazra2020} show that the inclusion of the additional mean-field $\alpha$-effect leads to complexities that may impact the memory and predictability of the model.

%\textit{Furthermore, BN18 did not give the cycle period of their dynamo model. According to Figure 2 of \cite{Passos2014}, who provided the standard dynamo model used by BN18, the cycle period is about 12 years. The longer period might be the reason that although they did not extend or compress the observed cycles, the cycle maximum phases of observations and simulations still  roughly match with each other based on their Figure 4. The predictive model can give the strength and timing of the cycle 25 maximum phase.}

\subsubsection{LCL19}

LCL19 initiate a new generation of physics-based solar cycle prediction by a coupled SFT-FTD simulation. The interior FTD module provides the synthetic active region emergences driving the SFT module, while the SFT provides the upper boundary condition that acts as the poloidal source term of the FTD. The predictive model consists of two modes of operation: a ``data-driven mode", using the active region database to insert active regions in the SFT module; and a ``dynamo mode", using the emergence function setting the probability and properties of active region emergence in the SFT module. A forecast is produced by running the simulation in data-driven mode until November 2017, and switching to dynamo mode to generate an ensemble of simulations with statistically independent realizations of active regions emergences. The forecast is constructed from the statistical characterization of this ensemble. The model can give the possible profiles of cycle 25, including amplitude, rising and decline phases, and northern and southern asymmetry.

Three mechanisms and three sets of corresponding parameters are involved in LCL19. The first is the surface flux transport (SFT) mechanism. Since the poloidal field provides the seed to the future cycle(s), the SFT module is essential for the prediction. \cite{Cameron2013, Jiang2014, Nagy2017} have shown that the source parameters play a dominant role in the surface field evolution. In ``data-driven mode" for cycles 23 and 24, the data sources and reduction procedure are based on \cite{Yeates2008}. In ``dynamo mode", the parameters of the active region are based on statistical distributions constructed from the database assembled by \cite{Wang1989} for cycle 21. A genetic algorithm-based optimizer is applied to specify the values of 18 transport parameters and the source parameters \citep{Lemerle2015}. Some transport parameters actually could be constrained by observations. Although both AR datasets are determined from NSO synoptic maps, they are subject to different detection methods. Furthermore, NSO changed their observations from the KPVT (868.8nm line) to SOLIS/VSM (630.2nm) at CR2007 (2003 August). Some observational gaps are also presented in the original NSO synoptic maps. Active region parameters, e.g., tilt angle and latitude, are cycle-dependent \citep{Dasi2010,Jiao2021}, which are non-linear mechanisms to modulate the poloidal field generation \citep{Jiang2020, Karak2020}. The factors listed above might potentially contribute to the source of the model's systematic errors.

The second mechanism is about the flux transport in the convection zone. The flux transport process determines the magnetic memory, and hence affects the prediction result. As shown by Table 1 of \cite{Lemerle2017}, the ``dynamo mode" uses a considerably higher turbulent diffusivity $(\sim10^{12}\textrm{cm}^2\textrm{s}^{-1})$ than was possible before. The value is close to the estimation based on
the mixing-length theory, see \cite{Munoz-Jaramillo2011, Karak2016} for more discussions.  A rapid equatorward meridional flow is also used. Its peak value is about 6.6 $\textrm{m}\textrm{s}^{-1}$, which is significantly larger than the widely adopted value 2-3 $\textrm{m}\textrm{s}^{-1}$. It takes only a few years for the rapid equatorward meridional flow to carry the polar flux into the tachocline to generate the toroidal field, as shown by Figure 7a of \cite{Lemerle2017}. The memory of the dynamo model is one cycle.

The third mechanism is destabilization and emergence of magnetic flux tubes. This is a poorly understood component of the solar cycle. LCL19 apply a surface flux deposition technique through masking of the deep-seated toroidal field based on a semi-empirical emergence function. This part corresponds to the connection between the SFT and FTD modules and provides the source of the SFT module. Except for the source term discussed in the first mechanism, the threshold value ($B^{*}$) above which active region emergence takes place and dynamo number ($K$) are two key parameters affecting the model. Their values might potentially be a source of systematic errors in the prediction model.

%\iffalse

%\fi

\section{A comparison of SFT initial conditions on prediction results}
\label{sec:detailedCom}

\begin{figure}[t]
  \centering
  \includegraphics[scale=0.6]{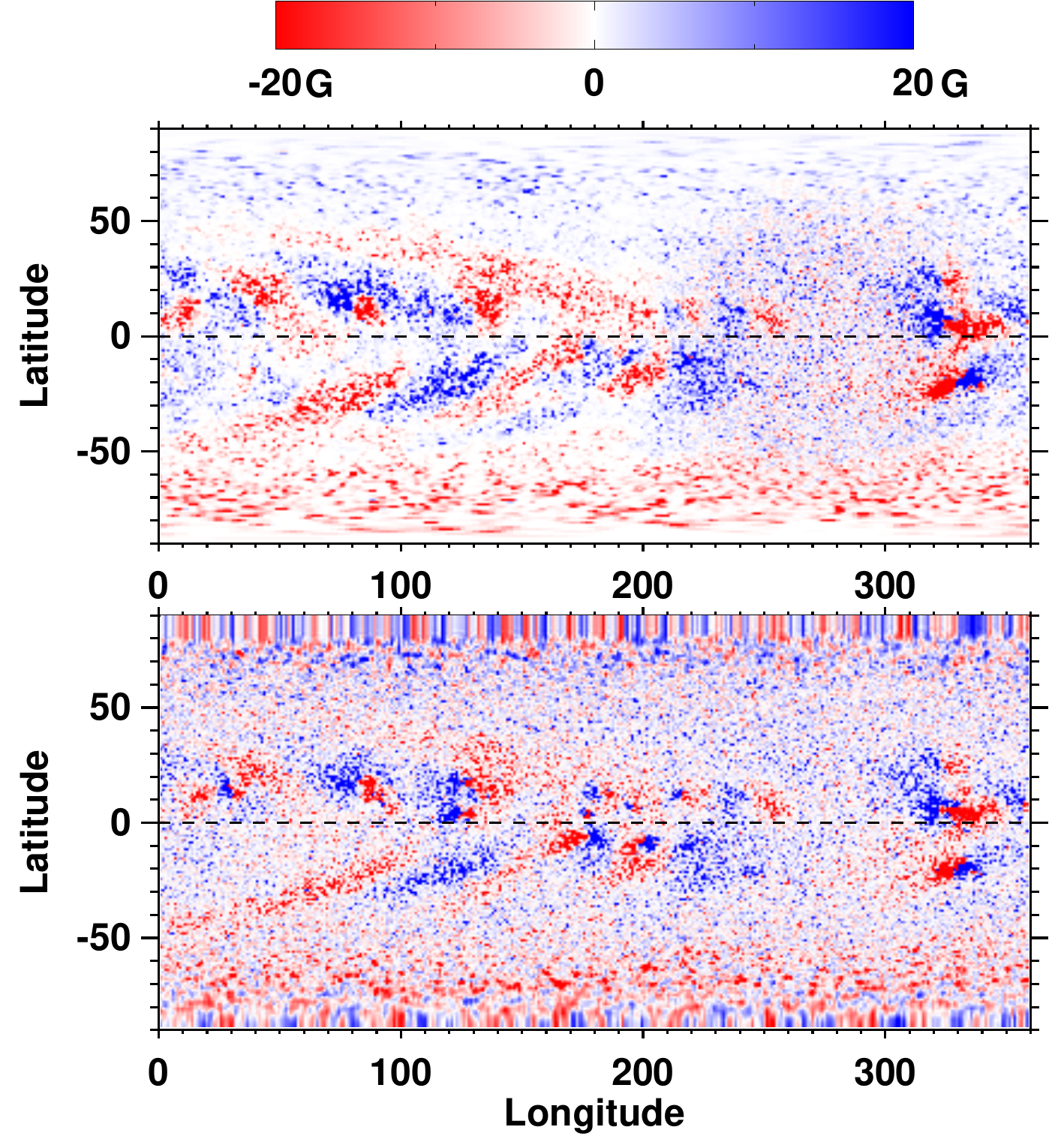}
  \caption{Comparisons of magnetograms at CR2172 (December 25, 2015-January 21, 2016) used by different SFT-based solar cycle predictions. Upper panel: synchronic map provided by Lisa Upton. Lower panel: synoptic map from SDO/HMI with interpolation to be equal space in longitude and latitude.}
  \label{fig:CompMag}
  \end{figure}

  \begin{figure}[h!]
  \centering
  \includegraphics[scale=0.6]{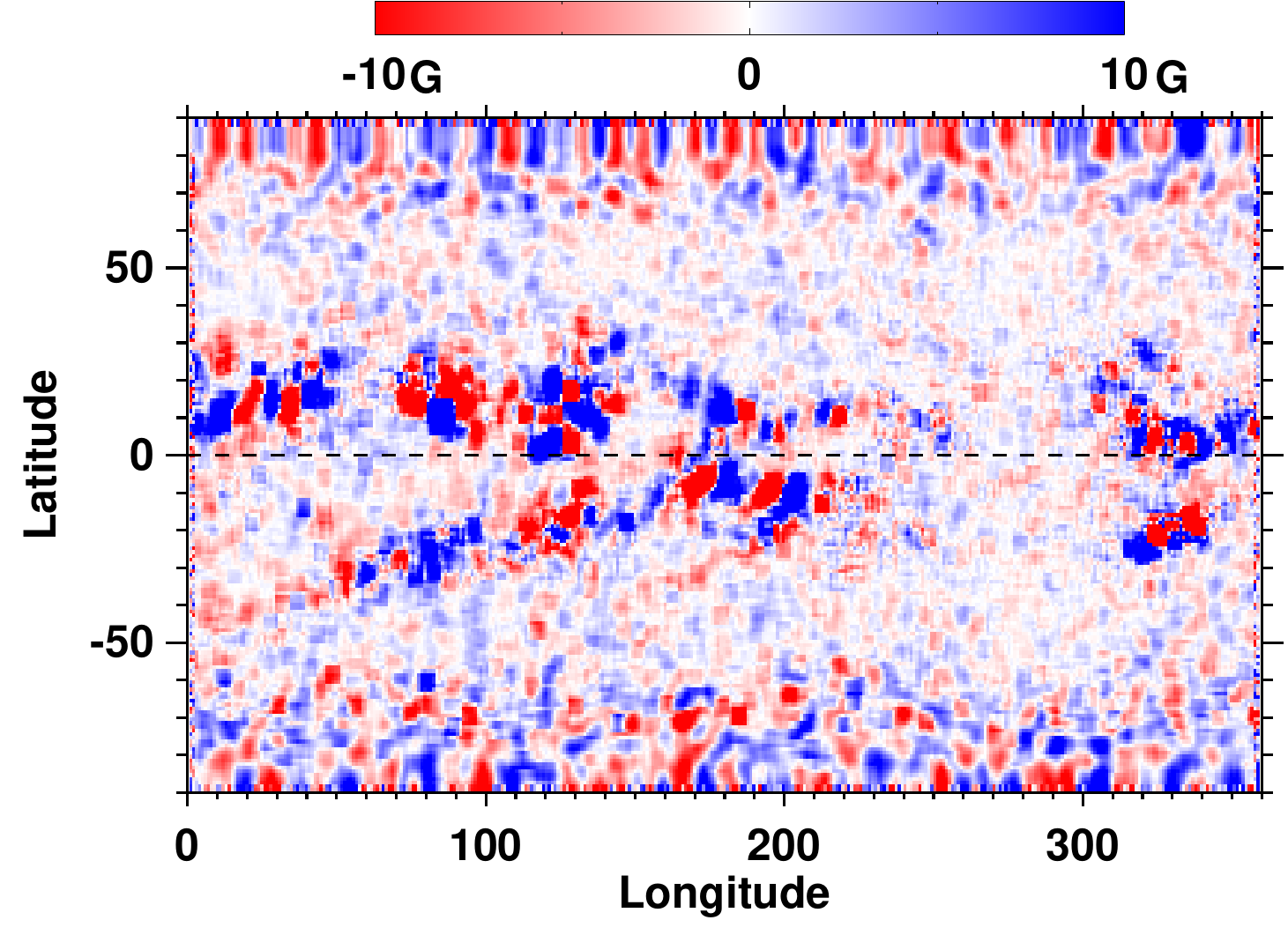}
  \caption{Difference between the maps shown in the upper and lower panels of Figure \ref{fig:CompMag}. }
  \label{fig:diffMag}
  \end{figure}

  \begin{figure}[h]
  \centering
  \includegraphics[scale=0.6]{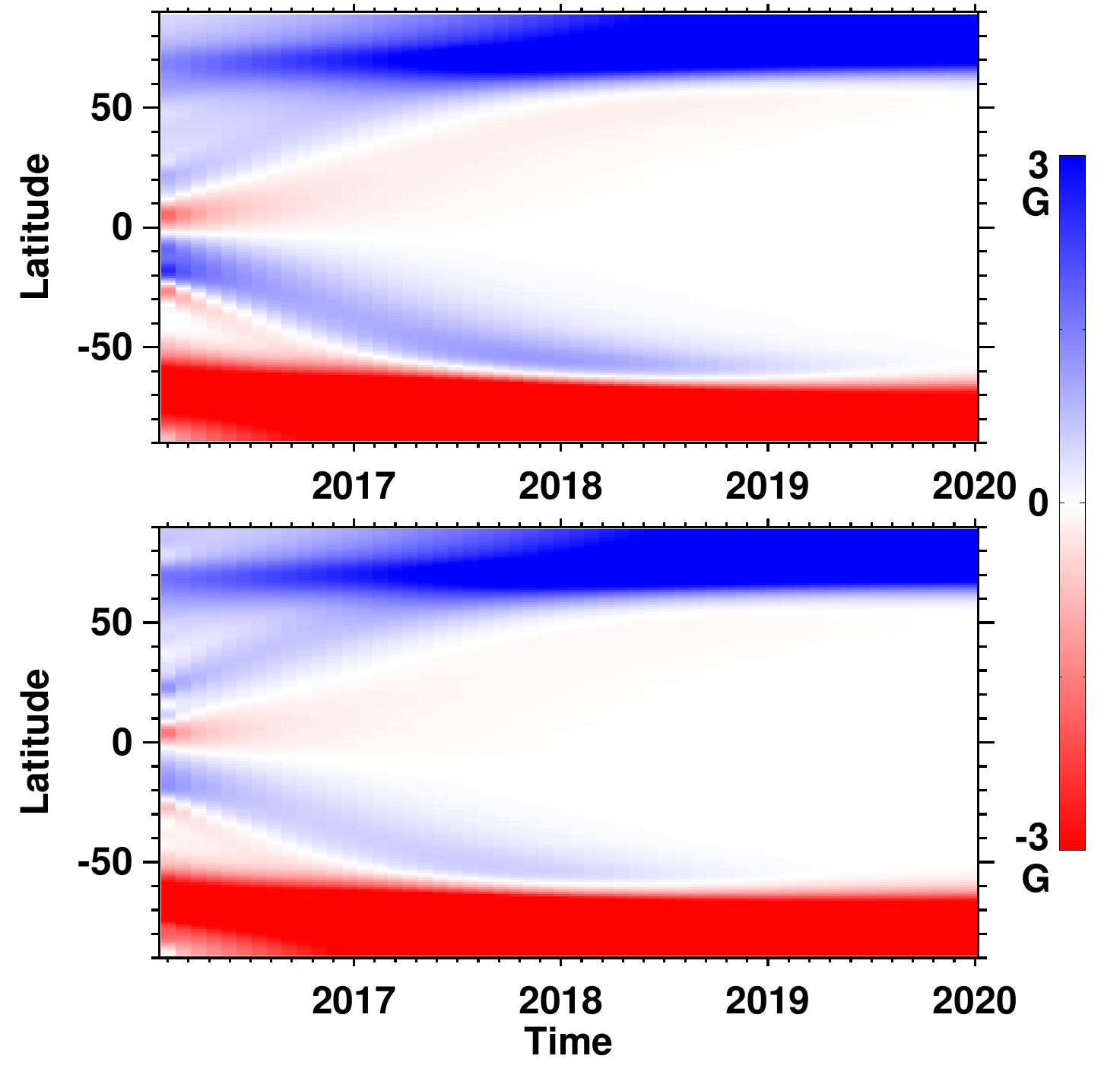}
  \caption{Time evolution of the longitudinally averaged magnetograms derived based on the SFT simulations using the synchronic map (upper panel) and synoptic map (lower panel) of CR2172 as the initial conditions, respectively. No flux emergence during the simulation time period, i.e., 2016-2020, is included.}
  \label{fig:MagBtf}
  \end{figure}

  \begin{figure}[h]
  \centering
  \includegraphics[scale=0.7]{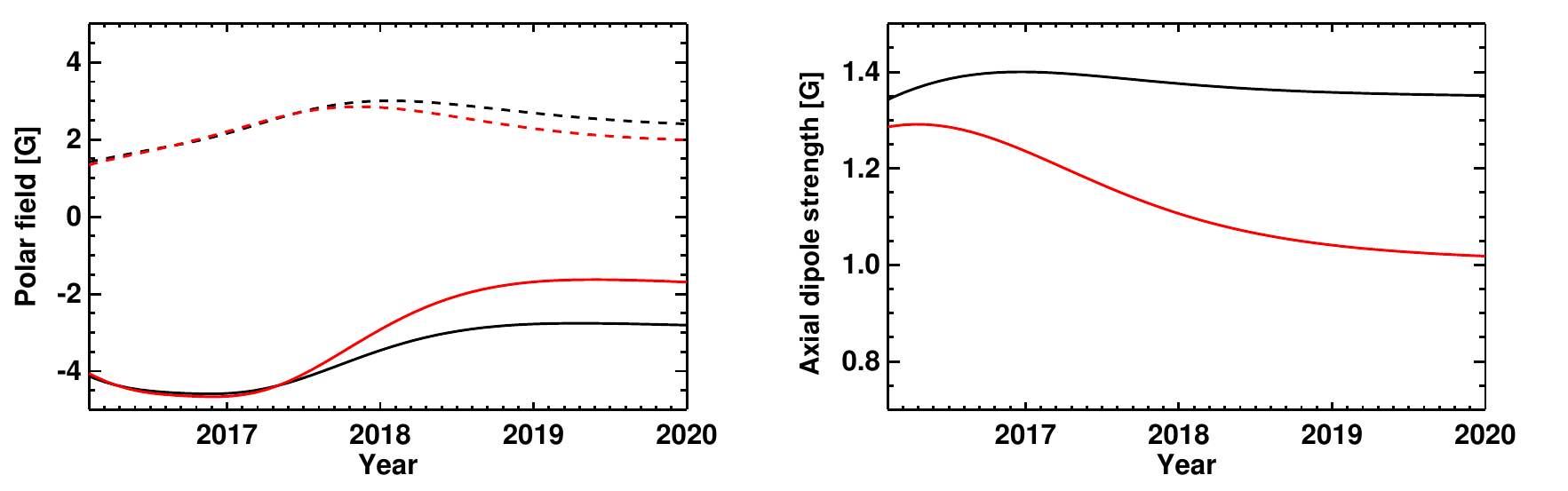}
  \caption{Comparisons of the time evolution of the polar field (left panel, southern/northern pole in solid/dashed curves) and  axial dipole strength (right panel) from the two SFT simulations using the synchronic map (red curves) and synoptic map (black curves) of CR2172 as the initial conditions, respectively.}
  \label{fig:DM_PF_diff}
  \end{figure}

%*****
%They assimilate data from magnetograms on the Sun��s near side but the field evolution on the far side is produced purely by the surface flux transport.
%The surface flux transport advection equation was solved with explicit time differencing to produce magnetic flux maps
%of the entire Sun with a cadence of 15 minutes. (These maps are referred to as synchronic maps since they represent
%the Sun��s magnetic field at a moment in time.)
%****

%\iffalse
In the previous section, we have mentioned that CJS16 and HU16 use similar physical methods to predict the cycle 25 strength. However, they differ not only in their exact predicted values but also in the predicted temporal variations of the polar fields and their asymmetry. The different methods to deal with the BMR emergence have been presented there. Here we give a detailed analysis of the effect of SFT initial conditions on the prediction result.

What CJS16 used were the observed synoptic magnetograms provided by instruments, e.g., SOHO/MDI or SDO/HMI. What HU16 used is the synchronic maps constructed by their advective flux transport model \citep{Upton2014, Ugarte2015}. Here, we utilize the maps of the same CR as the initial fields of SFT simulations. We will not include any BMR emergence during the SFT simulations. And we will use the same SFT code, which is the one used by CJS16 and is originally from \cite{Baumann2004}. Thus, we can concentrate on analyzing the effects of the different initial fields on the polar field evolution. 

We choose a specific timing, i.e., CR2172 (December 25, 2015-January 21, 2016) as the reference to compare. The synchronic map of CR2172 was produced by courtesy of Lisa Upton. The map has 1024 points in Carrington longitude by 512 points in latitude. The numerical resolution of the SFT code is 361 in longitude and 181 in latitude. Hence we use the IDL function CONGRID to shrink the size into 361 by 181. Its distribution is shown in the upper panel of Figure \ref{fig:CompMag}. It will be used as the first initial condition of the SFT simulation. HMI CR2172 synoptic map, i.e. ``hmi.Synoptic\_Mr\_720s.2172.synopMr.fits'', has 3600 points in Carrington longitude by 1440 points equally spaced in sine latitude. At a given longitude the points having strength above 30 G above $\pm75^\circ$ or having NaN values are regarded as the false points. Other points are taken as the true points. We use the average value of the 5 true points closest to the poles to fill the false points. Then we use the IDL function SMOOTH with a width value of 7 to smooth the map. After that, we interpolate the map onto a grid uniform in latitude. Finally, we use the CONGRID function to compress it into 361 points in longitude by 181 points in latitude. Its distribution is shown in the lower panel of Figure \ref{fig:CompMag}. The final map is the second initial condition of our SFT simulation.

At first sight, the upper and lower panels of Figure \ref{fig:CompMag} look similar. The differences between the two maps are shown in Figure \ref{fig:diffMag}. The differences concentrate in two belts. One is near the poles. Our concern here is the variation of the dipole moment, which is proportional to $\sin\theta\cos\theta$. The values near the poles have minor effects on the result. The other difference is around the activity belt, especially within $\pm20^{\circ}$ latitudes.

In the following we do SFT simulations with the code used by CJS16 to show the effects of the initial magnetograms on the dipole strength evolution. We take the same profiles of the transport parameters, i.e., turbulent diffusion, meridional flow, and differential rotation in the SFT model as CJS16 did. The two maps shown in the upper and lower panels of Figure \ref{fig:CompMag} are taken as the initial fields of the two SFT simulations, respectively. We run each SFT simulation for 4 years, that is, until the end of cycle 24.

Figure \ref{fig:MagBtf} shows the time evolution of longitudinally averaged solar surface magnetic field $\langle B_r\rangle$. Results from the synchronic map (upper panel) show the presence of surges of leading polarity propagating towards the poles, the surge on the southern hemisphere being particularly strong. It is also apparent that the surges in the upper panel (synchronic map case) are much stronger than in the lower panel (synoptic map case). These surges must have the effect of reducing the polar fields especially in the South, and especially in the synchronic map case.

This expectation is confirmed by the left panel of Figure \ref{fig:DM_PF_diff}, which shows the comparison of the polar field evolution for the two cases. The polar field is defined as the average of $\langle B_r\rangle$ over 60$^\circ$ and 75$^\circ$ in latitudes. Although there are differences in the flux distributions near the pole regions between the two input maps, as shown in Figure \ref{fig:diffMag}, the difference almost has no effect on the polar field evolution since it is defined lower than $\pm75^\circ$. The right-hand panel shows the time evolution of the corresponding axial dipole strength calculated based on Eq.(\ref{eqn:dm_t}). At the beginning of the simulations, the two maps show similar polar field strengths. The dipole moment from the synchronic map (red curve) is slightly lower than that from the HMI synoptic map (black curve) due to different initial magnetic flux distributions. The surges originating near the equator reach the polar areas in about one year in the Southern hemisphere and about 1.5 years in the northern hemisphere. Due to the different strengths of these surges in the two cases considered, the two curves representing polar fields start to deviate from each other. At the end of cycle 24, the southern polar field from the synchronic map (red solid curve) is strikingly weaker than that from the HMI synoptic map (black solid curve). Similarly, the northern polar field from the synchronic map (red dashed curve) shows a stronger decrease than that from the HMI synoptic map (black dashed curve). Eventually the two initial maps, which seem similar, give rise to starkly different polar field strengths and dipole moments at the end of the cycle, the dipole moments differing by $\sim 30\,${\%}.

The experiment reveals that the initial conditions play a dominant role in the physics-based predictions, in which the axial dipole strength at cycle minimum determines the subsequent cycle strength. \cite{Wang2021} suggest the mathematical deduction and algebraic quantification of an AR's contribution to the final dipole moment, i.e., the dipole moment at the cycle minimum. The deduction and algebraic quantification can be extended to the synoptic and synchronic magnetograms since the SFT model is linear. The contribution of an arbitrary magnetogram to the final dipole moment is a linear addition of the contribution of its smaller parts. The different fluxes around the equator significantly affect the predicted axial dipole strength at the cycle minimum. Since the prediction is sensitive to initial magnetograms, the measurement errors in magnetograms bring uncertainty to the prediction. \cite{Cameron2016b, Jiang2018b} are the only prediction models that include effects of the measurement errors in magnetograms. But they only analyze the effects of the net flux on the uncertainty. Effects of other measurement errors in magnetograms, e.g., center-to-limb noise variation, deserve to be further investigated.

%Partly agree with Petrovay (2020): these first attempts at physics-based solar cycle prediction are now generally seen as precursor methods in disguise. If the interior flux transport is correct, FTD-based predictions should be consistent with precursor prediction and with SFT-based prediction.
%possible problem in observed polar field, but the relative amplitudes are our concern.

\section{Conclusions and discussion}\label{s:Conclusion}
%\iffalse
In the paper we have compared 7 physics-based prediction models of cycle 25. Four of them rely on SFT models to obtain the axial dipole strength at the cycle 24/25 minimum. The differences among predictions of this type concentrate on three aspects. The first one is the choice of the initial conditions for the SFT models, e.g., synoptic maps from SDO/HMI observations or synchronic maps synthesized by models. The different flux distributions, especially near the equator, significantly affect the predicted axial dipole strength at cycle minimum. The second difference is the way to represent future flux emergence up to the next minimum: no flux emergence, random realizations based on empirically derived statistical relations, or using active regions from cycle 14. The difference in treating flux emergence affects both the amplitude and uncertainty range of the cycle 25 prediction. The third difference is how to obtain a cycle 25 prediction based on the axial dipole strength at the cycle 24 minimum. One way is only based on the correlation between the polar field at cycle minimum and the subsequent cycle strength. The other way is based on both the correlation and the empirical relations between cycle shapes and cycle amplitudes. The third difference determines whether only the cycle amplitude or the whole cycle shape can be predicted. Two further models assimilate the surface poloidal field at cycle minimum into FTD models to get the cycle 25 prediction. These two models differ in the assimilated datasets. One is from direct magnetic field observations, which just cover 4 entire cycles. The other is from SFT simulations for 12 cycles. Finally, one model makes the first attempt to apply the coupled surface poloidal field evolution with an FTD to the solar cycle prediction. The axial dipole strength or polar field at cycle minimum is essential to all physics-based predictions.

Reviews on solar cycle predictions usually place emphasis on the comparisons of prediction results. In contrast, in this review we do not address the predicted values for the amplitude of cycle 25 from different models for the following reasons. First, other available reviews, e.g., Figures 4 and 5 of \cite{Nandy2021} have given a complete comparison of the predicted cycle 25 strengths. Secondly, and more importantly, here we focus on physics-based prediction models. There are still many open questions about physical mechanisms underlying the solar cycle and solar interior dynamics, see Section 8 of \cite{Charbonneau2020} for more details. While all 7 physics-based predictions considered are based on the BL mechanism, they differ in many of their details, and some parameters are often optimized to fit the observations. Furthermore, even if the BL mechanism captures the essence of the solar cycle, we do not have the perfect surface magnetic measurements. The imperfect magnetograms as the assimilated data into the prediction models must bring uncertainties, which are not well evaluated. Our preliminary test in Section \ref{sec:detailedCom} has shown the large effects of the observations on predictions. Other well-known ingredients of the solar dynamo, e.g., the meridional flow at the surface are still controversial. This is demonstrated in Table \ref{table:1}. Hence a successful prediction does not mean that the physics-based prediction model is ideal and captures all the physics of the solar cycle. On the other hand, a failed model may still valuably contribute to understanding the solar cycle.

In summary, the physics-based prediction of a solar cycle is still in its infant stage. It provides an effective way to verify our understanding of the solar cycle. The value of a prediction model cannot be simply judged by the consistency of a prediction result with the observation. But if cycle 25 is confirmed as a medium cycle in a few years, it will add a further piece of evidence that the polar field determines the subsequent cycle and that the BL mechanism is at the heart of the solar cycle.

%\fi

~\\
\textbf{Acknowledgments.} We thank the two anonymous referees for carefully checking the manuscript. We further thank Lisa Upton for offering their synchronic map of CR2172 and Robert Cameron for always valuable discussions. We acknowledge the Solar Cycle 25 Prediction workshop held by Nagoya University in November 2017 and the International Space Science Institute Teams 474. J.J. and Z.B.Z. are supported by the National Natural Science Foundation of China (grant Nos. 11873023 and 12173005). K.P. is supported by the Hungarian National RD\&I Fund (grant No. NKFI K-128384) and by the EU H2020 grant No. 955620. The SDO/HMI data are courtesy of NASA and the SDO/HMI team. Wilcox Solar Observatory data used in this study was obtained via the web site http://wso.stanford.edu courtesy of J.T. Hoeksema. The sunspot records are courtesy of WDC-SILSO, Royal Observatory of Belgium, Brussels.

~\\
\textbf{Disclosure of Potential Conflicts of Interest} The authors declare that there are no conflicts of interest.

%%%%%%%%%%%%%%%%%%%%%%%%%%%%%%%%%%%%%%%%%%%%%%%%%%%%%%%%%%%%%%%%%%%%%%%%%%%
%% The Appendices part is started with the command \appendix;
%% appendix sections are then done as normal sections

%
%\bibliographystyle{cas-model2-names}
%\bibliography{Reference}

\end{document}